\documentclass{PoS}

\usepackage{epsfig,multicol,multirow}

\newcommand{\be}{\begin{equation}}
\newcommand{\ee}{\end{equation}}
\newcommand{\bea}{\begin{eqnarray}}
\newcommand{\eea}{\end{eqnarray}}

\title{Dynamical overlap fermion simulations with a preconditioned Hybrid Monte Carlo force
\thanks{{Preprint ~ HU-EP-06/23, BI-TP 2006/28, SFB/CPP-06-39.}} }

\ShortTitle{Dynamical Overlap with Preconditioned Force}

\author{\speaker{Jan Volkholz} \ and \ Wolfgang Bietenholz \\
  Institut f\"{u}r Physik, Humboldt-Universit\"{a}t zu Berlin \\
  Newtonstr.\ 15, D-12489 Berlin, Germany \\
  E-mail: \email{volkholz@physik.hu-berlin.de, bietenho@physik.hu-berlin.de}}

\author{Stanislav Shcheredin \\
Fakult\"{a}t f\"{u}r Physik, Universit\"{a}t Bielefeld \\
D-33615 Bielefeld, Germany \\
E-mail: \email{shchered@physik.hu-berlin.de}}

\abstract{We present simulation results for the 2-flavour Schwinger 
model with dynamical Ginsparg-Wilson fermions.
Our Dirac operator is constructed by inserting
an approximately chiral hypercube operator
into the overlap formula, which yields the overlap hypercube  
operator. Due to the similarity with the hypercubic kernel,  
a low polynomial of this kernel can be used as a numerically 
cheap way to evaluate the fermionic part of the Hybrid Monte Carlo force. 
We verify algorithmic requirements like area conservation
and reversibility, and we discuss the viability of this approach in   
view of the acceptance rate. Next we confirm a high level of locality 
for this formulation.
Finally  we evaluate the chiral condensate at light fermion masses,
based on the density of low lying Dirac 
eigenvalues in different topological sectors. The results
represent one of the first measurements with dynamical
overlap fermions, and they
agree very well with analytic predictions.}

\FullConference{XXIVth International Symposium on Lattice Field Theory\\
                July 23-28, 2006\\
                Tucson, Arizona, USA}

\begin{document}

\section{Motivation}

\vspace*{-2mm}
In 1998 a neat possibility was found to preserve
chiral symmetry (in a modified form) on the lattice \cite{Has,Neu,ML}. 
It has been used extensively in quenched QCD, but due to its
computational demands the applications of {\em dynamical} chiral fermions 
are still in an early stage.\footnote{Recent status reports on dynamical
overlap fermion simulations in 
QCD are given for instance in Refs.\ \cite{Cundy}.}
Hence it is strongly motivated to develop suitable
algorithmic tools, in order to arrive --- within a few years --- at results 
that can be confronted with the light hadron phenomenology.

Present chiral QCD simulations are restricted to coarse lattices;
in particular thermodynamic studies typically use 
$N_t=4$ and $a \approx 0.28 ~ {\rm fm}$.
On such lattices the standard overlap operator is non-local.
Locality improves, however,
if we replace the kernel by a truncated perfect hypercube operator
\cite{EPJC}. In quenched QCD, the locality of the resulting
{\em overlap hypercube fermion} (overlap-HF) operator persists on 
rather coarse lattices \cite{QCD}. 
At present, dynamical hypercube fermion (HF) simulations for
QCD are under investigation \cite{Stani06}; they will indicate
how far the above property still holds beyond the quenched approximation. 
Thermodynamic tests show already that the cutoff effects 
for the HF are pushed to high energy \cite{thermo}.

In this work, we explore the feasibility of dynamical overlap-HF
simulations with an algorithm which is peculiar to this type
of Ginsparg-Wilson fermions. Our testing ground is the Schwinger model 
(2d QED), a popular toy model with certain features similar to 
QCD.\footnote{Earliest efforts to 
simulate the Schwinger model with 
dynamical overlap fermions were reported in Ref.\ \cite{Dubna}.}
Qualitative differences from QCD are the
super-renormalisability of the Schwinger model and the absence
of spontaneous chiral symmetry breaking.
We consider the case of two degenerate flavours.

\vspace*{-3mm}
\section{The overlap hypercube fermion}
\vspace*{-2mm}

The Ginsparg-Wilson Relation (GWR) 
is a criterion for a lattice 
modified, exact chiral symmetry \cite{ML}, which was discovered by 
studying the properties 
of perfect and classically perfect \cite{Has} lattice fermions. 
Since those formulations involve couplings over an infinite range, 
a truncation is needed, which
distorts the perfect symmetry and scaling properties to some extent.
For the free, optimally local, perfect fermion \cite{WBUJW} the truncation to a 
unit hypercube 
preserves excellent scaling \cite{BBCW} and chirality \cite{EPJC}. 
It leads to the form 
$D_{{\rm HF},xy} = \rho_{\mu}(x-y) \gamma_{\mu} + \lambda (x-y)$,
i.e.\ a vector term plus a scalar term ($x,y$ are lattice sites).
In $d=2$ these terms involve only couplings to nearest neighbours and
across the plaquette diagonals. We gauge
$D_{\rm HF}$ by multiplying the compact link variables 
$U_{x,\mu} \in U(1)$
along the
shortest lattice paths connecting $x$ and $y$ (for the diagonal the
two shortest paths are averaged) \cite{WBIH}. Thus we arrive at the operator
$D_{{\rm HF}, xy}(U)$, which describes the HF.\footnote{We are using 
here the HF version which is denoted as CO-HF (chirally optimised 
hypercube fermion) in the Ref.\ \cite{WBIH}.
This is optimal for our algorithm to be described in Section 3.}

Since $D_{\rm HF}$ is $\gamma_{5}$-Hermitian, 
$D_{\rm HF}^{\dagger} = \gamma_{5} D_{\rm HF} \gamma_{5}$,
the exact chirality (which got lost in the truncation) can be
restored by inserting $D_{\rm HF}$ into the overlap formula \cite{Neu}, 
which yields the overlap-HF operator 
\vspace*{-4mm}
\be  \label{overlap} 
\vspace*{-4mm}
D_{\rm ovHF}(m) = \Big( 1 - \frac{m}{2} \Big) D_{\rm ovHF}^{(0)} + m \ ,
\quad D_{\rm ovHF}^{(0)} = 1 + \gamma_{5} \frac{H_{\rm HF}}
{\sqrt{H_{\rm HF}^{2}}} \ , \quad H_{\rm HF} = \gamma_{5} (D_{\rm HF}-1) \ .
\ee
$H_{\rm HF}$ is Hermitian and $D_{\rm ovHF}^{(0)}$ fulfils the
GWR in its simplest form, $\{ D_{\rm ovHF}^{(0)} , \gamma_{5} \} = 
D_{\rm ovHF}^{(0)} \gamma_{5} D_{\rm ovHF}^{(0)}$ .
In practice we evaluate this operator by means of Chebyshev polynomials
--- after projecting out the lowest two modes of $H_{\rm HF}^{2}$, 
which are treated separately. The polynomial approximation
was driven to some absolute accuracy of $\varepsilon$ (see below), so 
we deal with $D_{{\rm ovHF}, \varepsilon}$.

Compared to H.\ Neuberger's standard overlap operator $D_{\rm N}$ \cite{Neu},
we replace the Wilson kernel $D_{\rm W}$ by $D_{\rm HF}$ \cite{EPJC}. 
Since the latter is an approximate solution to the GWR already, its 
transition $D_{\rm HF} \to D_{\rm ovHF}$ is only a modest
chiral correction (in contrast to the transition 
$D_{\rm W} \to D_{\rm N}$). 
This pro\-perty is illustrated in Fig.\ \ref{specfig},
which compares the spectra of $D_{\rm HF}$ and $D_{\rm ovHF}$ for a typical 
configuration at $m=0.03$ and $\beta =5$ 
on a $16\times 16$ lattice.\footnote{Throughout this work
we use the Wilson plaquette gauge action.}
\FIGURE{
  \centering
 \includegraphics[angle=270,width=.48\linewidth]{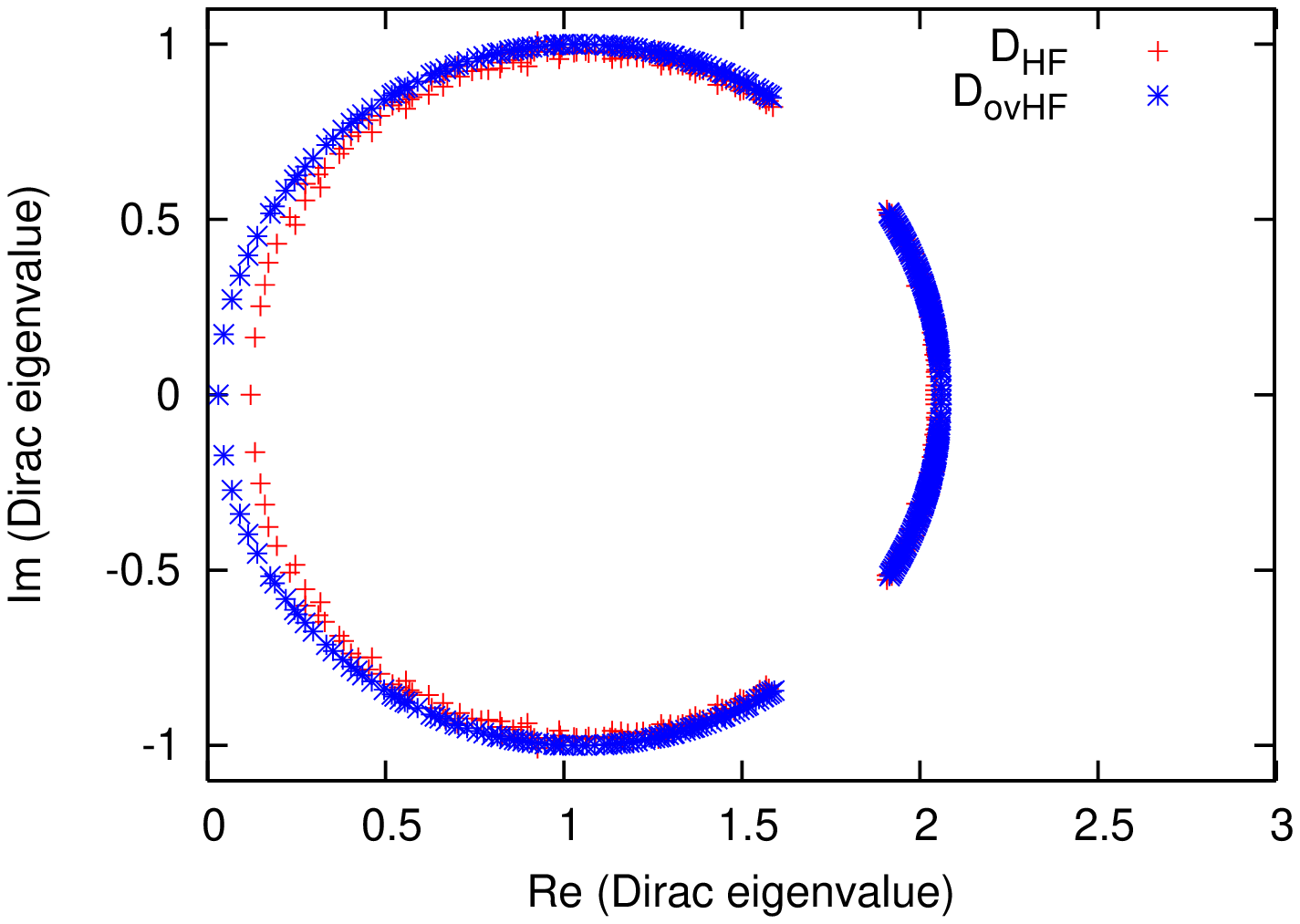}
  \caption{The spectra for $D_{\rm HF}$ and for $D_{\rm ovHF}$ 
(up to a gap)
at $m=0.03$, in a typical, dynamical configuration at $\beta =5$ on a
$16 \times 16$ lattice. Since the spectra are similar,
$D_{\rm HF}$ is a good approximation to $D_{\rm ovHF}$, and
therefore approximately chiral. \vspace*{-2mm}}
\label{specfig}
}

Due to its perfect action background, we expect for $D_{\rm HF}$
also a good approximate rotation symmetry and scaling behaviour, which
is then likely to be inherited by $D_{\rm ovHF}$ thanks to the relation
$D_{\rm ovHF} \approx D_{\rm HF}$. This relation further suggests a high
level of locality for $D_{\rm ovHF}$, since it deviates only a little
from the ultralocal operator $D_{\rm HF}$. These properties have been
confirmed before in a Schwinger model study with quenched configurations, 
where measurement entries were re-weighted with the fermion determinant 
\cite{WBIH}.

\vspace*{-3mm}
\section{A preconditioned Hybrid Monte Carlo force}
\vspace{-2mm}

In order to simulate such fermions dynamically, the standard 
Hybrid Monte Carlo (HMC) algorithm would use the fermionic force term
\bea
&& \hspace*{-1cm} \bar \psi \, Q^{-1}_{\rm ovHF} \Big( Q^{-1}_{\rm ovHF} 
\frac{\partial Q_{\rm ovHF}}{\partial A_{x,\mu}} + 
\frac{\partial Q_{\rm ovHF}}{\partial A_{x,\mu}} Q^{-1}_{\rm ovHF} 
\Big) Q^{-1}_{\rm ovHF} \, \psi \ , \label{HMCforce}
\eea
where $Q_{\rm ovHF} = \gamma_{5} D_{\rm ovHF}$ is the Hermitian overlap-HF 
operator, and $A_{x,\mu}$ are the non-compact gauge link variables.
However, this force term is computationally expensive, and in addition
conceptually problematic due to the discontinuous sign function
$H_{\rm HF} / \sqrt{ H_{\rm HF}^{2} } \, $ in $Q_{\rm ovHF}$, see eq.\ 
(\ref{overlap}). 

We render the force term continuous and computationally cheap
by inserting only approximate overlap 
operators in the term (\ref{HMCforce}).\footnote{Such a modified force
might also be helpful to achieve topological transitions more frequently,
but we have no data for comparison with the force (\ref{HMCforce}).}
For the external factors we apply an overlap-HF to a low precision 
$\varepsilon '$, and we use $H_{\rm HF}$ instead of $Q_{\rm ovHF}$ in the 
derivatives 
(although this could easily be extended to a low polynomial as well),
\be
\bar \psi \, Q^{-1}_{{\rm ovHF}, \varepsilon '} 
\Big( Q^{-1}_{{\rm ovHF}, \varepsilon '} 
\frac{\partial H_{\rm HF}}{\partial A_{x,\mu}} +
\frac{\partial H_{\rm HF}}{\partial A_{x,\mu}} 
Q^{-1}_{{\rm ovHF}, \varepsilon '}  
\Big) Q^{-1}_{{\rm ovHF}, \varepsilon '} \, \psi \ . 
\label{HFforce}
\ee
The Metropolis accept/reject step is still performed with the high precision
overlap operator $D_{{\rm ovHF}, \varepsilon}$. Hence the deviations
in the force are corrected, and the only point to worry about is the
acceptance rate.
A simplification, which reduces
$Q_{{\rm ovHF}, \varepsilon '}$ to $\gamma_{5} D_{\rm HF}$ ,
was originally proposed in Refs.\ \cite{CJNP} which reported a decreasing
acceptance rate for increasing volume 
(that work used the ``SO-HF'' of Ref.\ \cite{WBIH}).
However, 
it turned out to be highly profitable
--- and still cheap --- to correct the external factors
to a low precision. We chose
\vspace*{-2mm}
\be
\varepsilon ' = 0.005 \quad {\rm (force~term)} \ , \quad
\varepsilon = 10^{-16} \quad {\rm (Metropolis~step)} \ ,
\vspace*{-2mm}
\ee
which increases the acceptance rate by an order of magnitude
compared to the use of $D_{\rm HF}$ throughout the force term.
Note that the force we obtain in this way is not based on a
Hamiltonian dynamics, but the way we deviate from it (by 
proceeding from $\gamma_{5} D_{\rm HF}$ to $Q_{{\rm ovHF}, \varepsilon '}$)
does manifestly maintain the area conservation.

\vspace*{-3mm}
\section{Results for the acceptance rate, reversibility, locality and 
chiral condensate}
\vspace*{-2mm}

We performed production runs on a $16 \times 16$ lattice at $\beta =5$
with five
masses: $m = 0.03$, $0.06$, $0.09$, $0.12$ and $0.24$.
We applied the Sexton-Weingarten integration scheme \cite{SeWe} with a
partial $(\delta \tau )^{3}$ error cancellation (where $\delta \tau$
is the step size). The time scales for the fermionic vs. gauge force
had the ratio $1:5$, but we did not observe
a high sensitivity to 
this ratio. Our statistics of well thermalised configurations,
separated by 200 trajectories, is given in Table \ref{Sigmatab}.

Since the force (\ref{HFforce}) tends to push the trajectory a bit off
the hyper-surface of constant energy, we kept the trajectory length 
(between the Metropolis steps) short. We chose it as
$\ell = 1/8$, which is divided into 20 steps (i.e.\ $\delta \tau =
0.00625$); this turned out to be a good compromise in view of the acceptance
rate and the dynamics between the trajectory end-points. 
Fig.\ \ref{algofig} shows the acceptance
rate (on the left) as well as the total number of required 
conjugate gradient iterations per trajectory (on the right).
As usual, heavier fermions are easier to simulate. However, even down to
our lightest mass of $m = 0.003$ we obtained a useful acceptance 
rate $\approx 0.3$.
\FIGURE{
\vspace*{-8mm}
  \centering
  \includegraphics[angle=0,width=.5\linewidth]{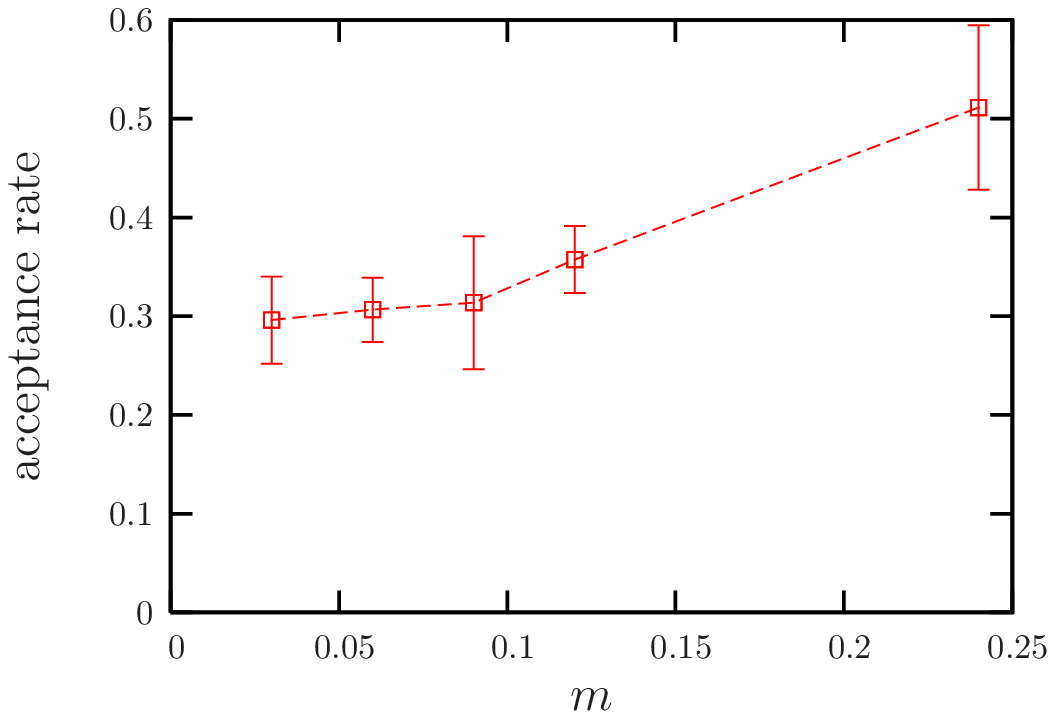}
\hspace*{-4mm}  
\includegraphics[angle=0,width=.5\linewidth]{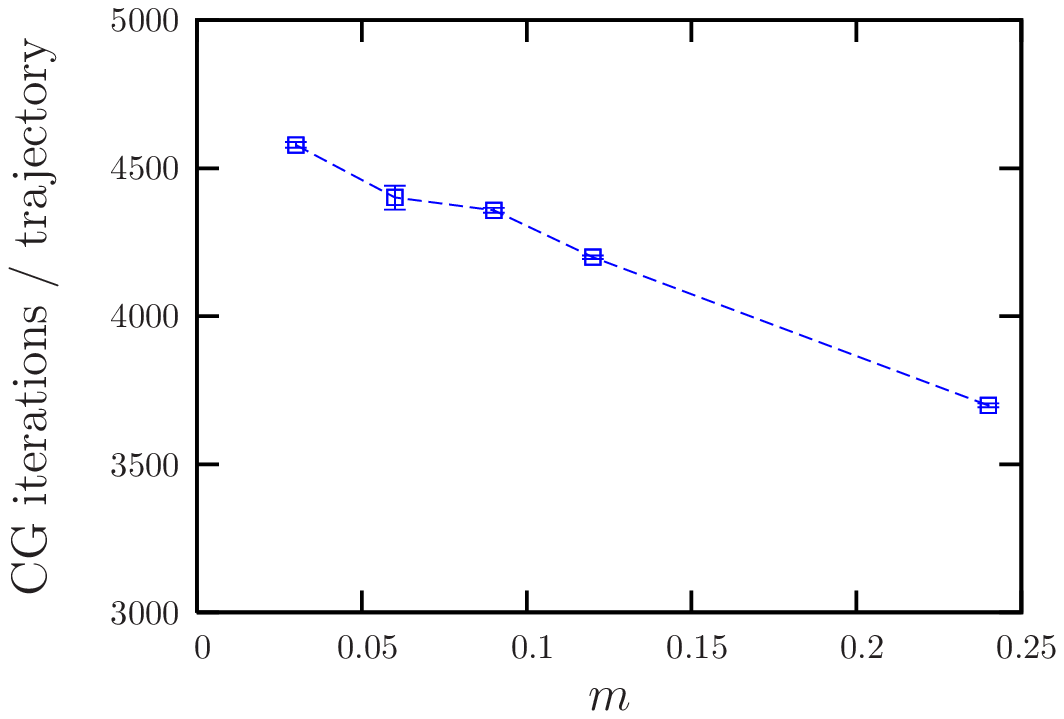}
 \caption{The acceptance rate (on the left), and the number of Conjugate
Gradient iterations per trajectory (on the right, including all operations) 
as a function of the 
fermion mass $m$, on a $16\times 16$ lattice at $\beta =5$ and trajectory length 
$\ell = 1/8 = 20 \cdot \delta \tau$.}
\label{algofig}
}

To study the quality of {\em reversibility,} we
moved forth and back with a variable number of steps, 
and measured the (absolute) shift of the gauge action, $| \Delta S_{G} |$.
Fig.\ \ref{revfig} shows our results for the 
precision of the reversibility, still at $\delta  \tau = 0.00625$,
for the masses $m=0.03$ and $0.12$ 
and $0.24$. 
The level of reversibility seems satisfactory. As we increase
the mass, it improves significantly only at $m = 0.24$,
as we also observed for $\delta \tau = 0.005$.
Our current results do not hint at any positive Lyapunov exponent,
though this cannot be considered conclusive yet.
\FIGURE{
  \centering
{\mbox{
  \includegraphics[angle=0,width=.45\linewidth]{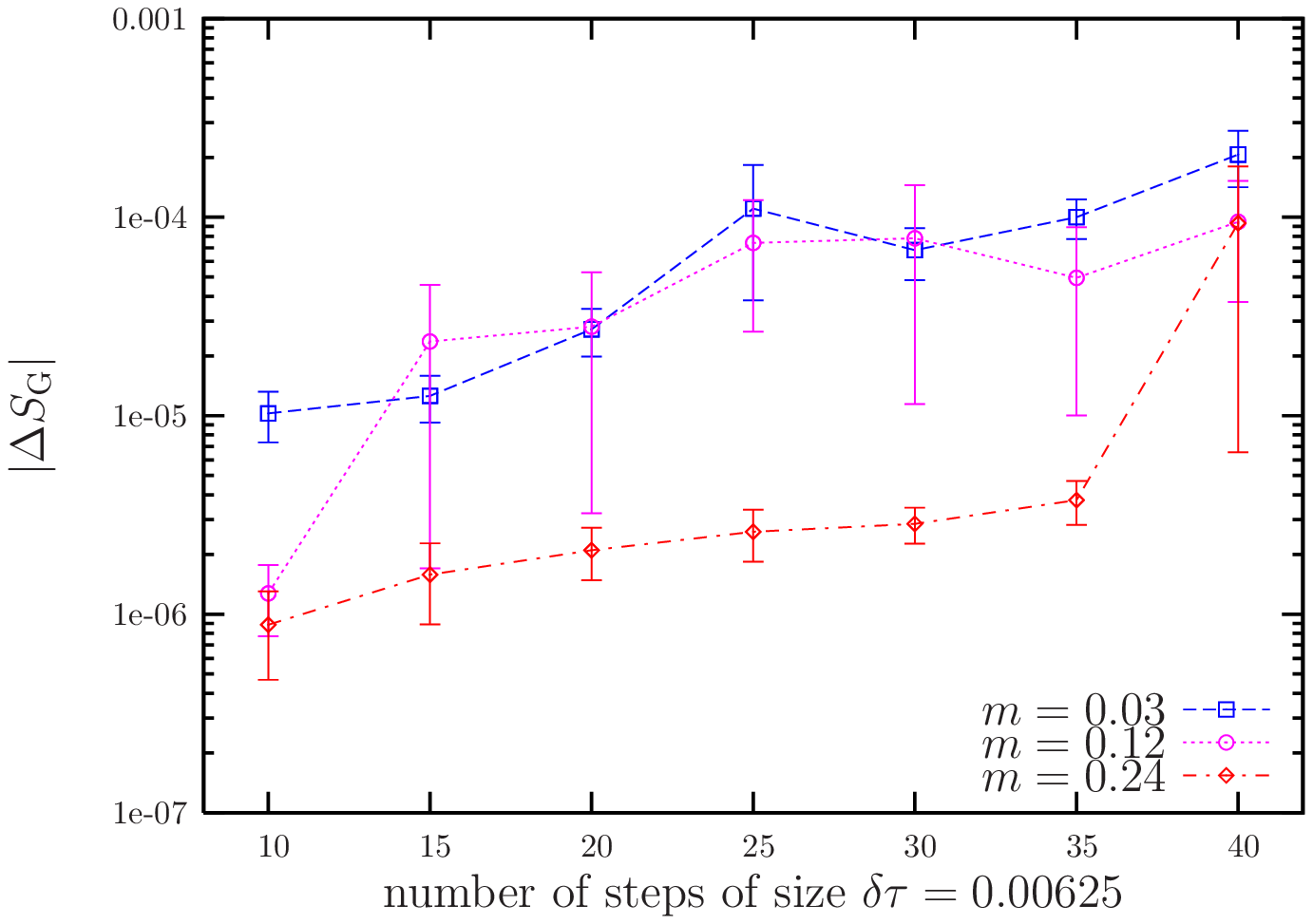}
}}
 \caption{The reversibility precision with respect to the gauge action
for a variable number of steps of length $\delta \tau =0.00625$.
We show the results for our lightest mass and the two heaviest masses. 
We do not see an obvious indication for a positive Lyapunov exponent.}
\label{revfig}
}


We tested the {\em locality} in the usual way \cite{HJL}, by applying 
$D_{{\rm ovHF}}$ on a unit source 
$\eta_{y} $ 
and measuring the decay of the function
\be  \label{floc}
f(r) = \ ^{\rm max}_{~ \ x} \ \Big\{ D_{{\rm ovHF},xy} \eta_{y} \ \Big| \ 
\sum_{\mu =1}^{2} | x_{\mu} - y_{\mu} |
= r \Big\} \ .
\ee
We first consider the free case and demonstrate that this decay is 
much faster for the overlap-HF operator than for the Neuberger
operator, see Fig.\ \ref{locfig} (on the left).
On the right we show
that the decay is still exponential for our dynamically generated 
configurations, which confirms the locality (and therefore
the sensibility) of our Dirac operator.
In the range that we considered, the mass has practically
no influence on this decay rate.
A previous quenched re-weighted study revealed that the overlap-HF
operator has a much higher degree of locality than the standard
overlap operator $D_{\rm N}$ \cite{WBIH}. 
This is observed here as well, since $D_{\rm ovHF}$ at $\beta =5$ is
still far more local than even the free $D_{\rm N}$.
We repeat that improved locality also holds for the 
overlap-HF in quenched QCD \cite{QCD}, and it enables the installation of
chiral fermions on coarser lattices than the use of $D_{\rm N}$.
\vspace*{-3mm}
\FIGURE{
  \centering
  \includegraphics[angle=270,width=.48\linewidth]{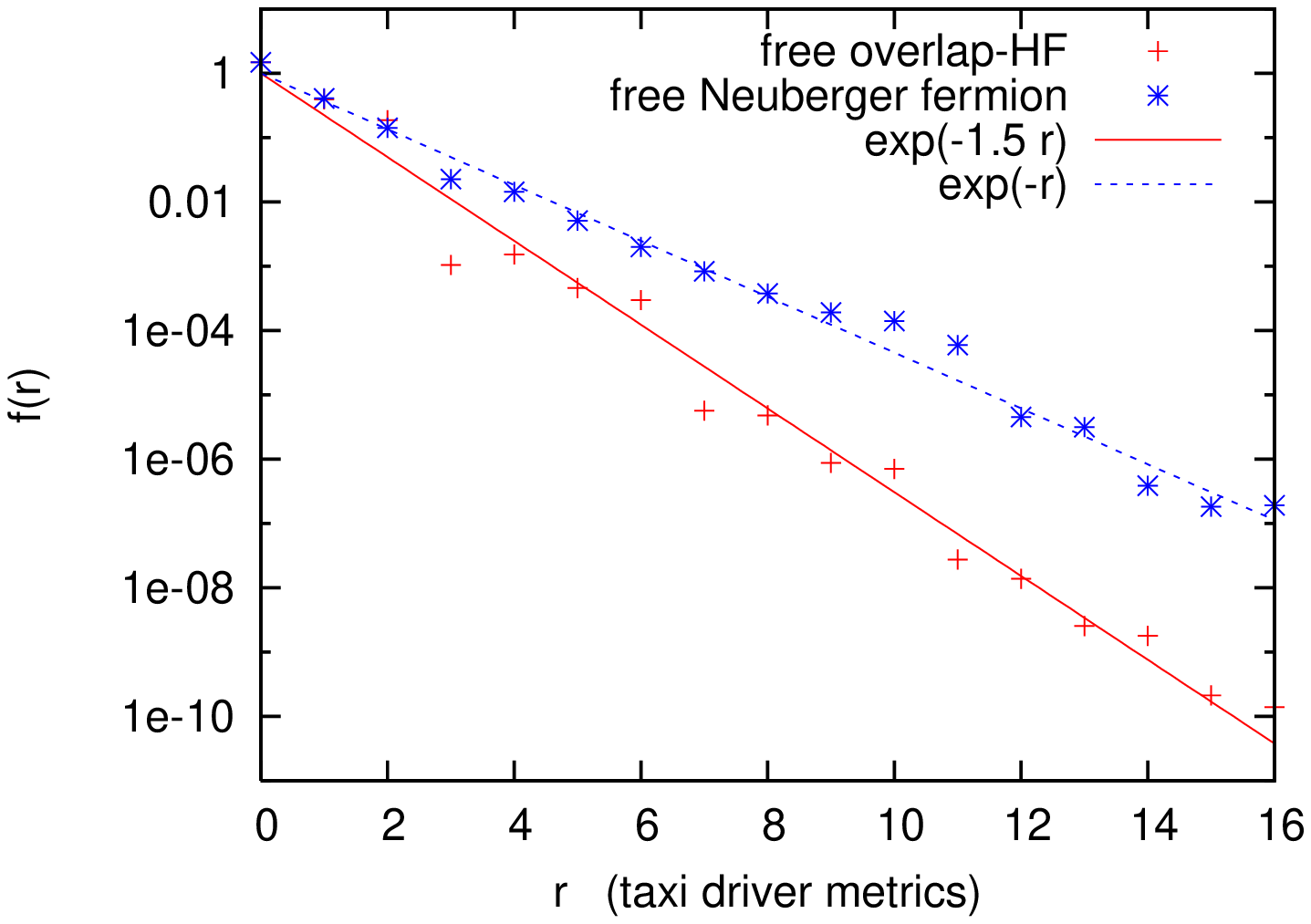}
\hspace*{-3mm}
  \includegraphics[angle=270,width=.48\linewidth]{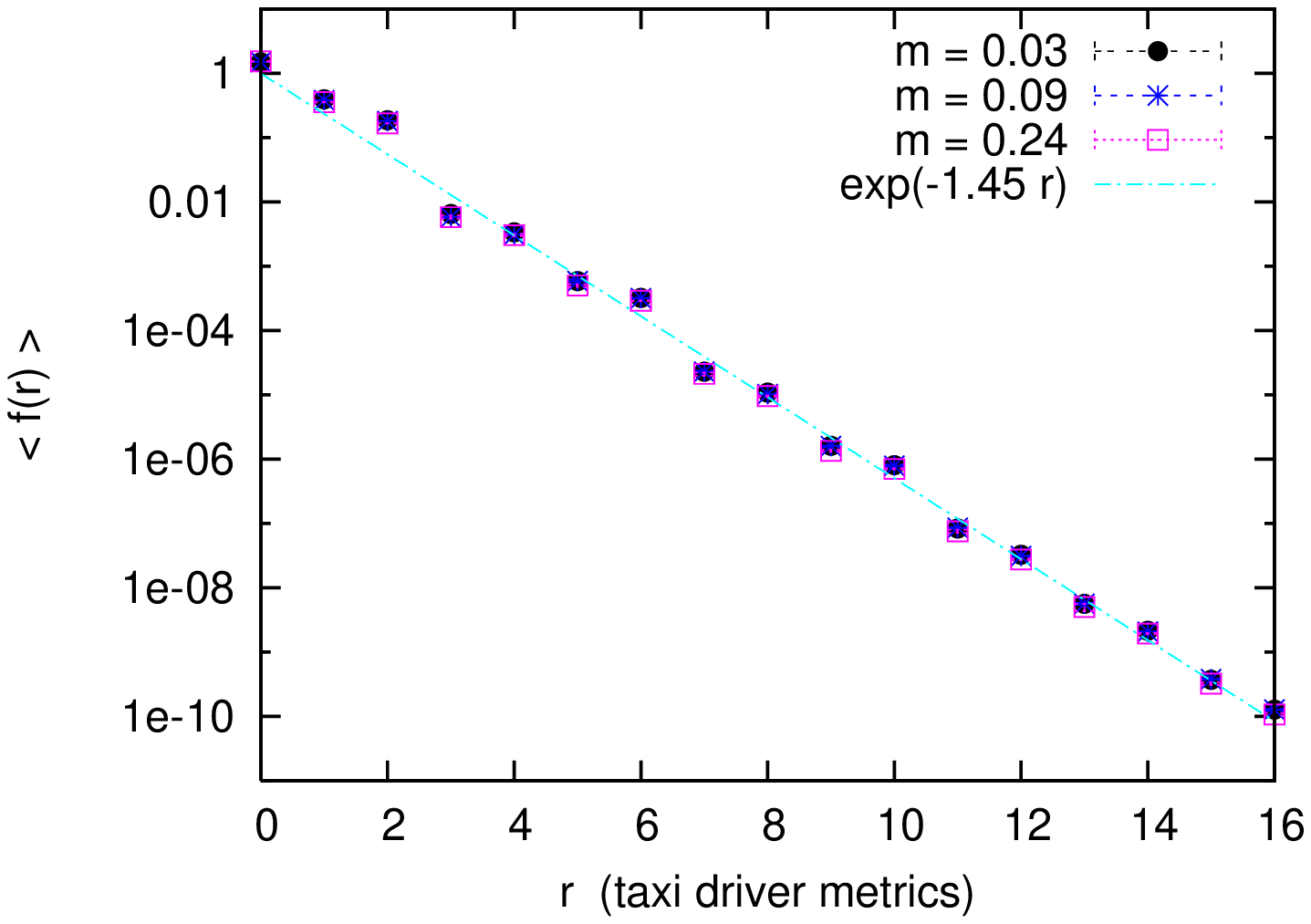}
\hspace*{-3mm}
 \caption{The locality of the overlap Dirac operators,
tested by the decay of the function (\ref{floc}),
against the taxi driver distance in lattice units.
On the left we compare our overlap-HF operator to the
standard Neuberger operator (with $H = \gamma_{5} (D_{\rm W} -1)$)
in the free case. The plot on the right shows the 
exponential decay of $\langle f(r) \rangle$ based on our
overlap-HF simulations with various fermion masses at $\beta =5$.}
\label{locfig}
}

\begin{table}
\centering
\begin{tabular}{|c||c|c|c|c|c|}
\hline
$m$ & 0.03 & 0.06 & 0.09 & 0.12 & 0.24 \\
\hline
\hline
$\#$ of confs.\ &               205 & 235 & 221 & 458 & 100 \\
\hline
$\#$ of confs.\ at $\nu=0$    &   0 &  79 &   0 & 325 & 1 \\
\hline
$\#$ of confs.\ at $|\nu |=1$ & 205 & 156 & 220 & 133 & 46 \\
\hline
$\#$ of top.\ transitions       & 0 &   1 &   2 &   3 & 5 \\
\hline
\hline
$\langle \lambda_{\nu =0} \rangle$ &
         & 0.129(3) &          & 0.108(2) & \\
\hline
$\langle \lambda_{| \nu | =1} \rangle$ &
0.171(2) & 0.173(2) & 0.171(2) & 0.165(3) & 0.169(4) \\
\hline
$\Sigma$ & & $0.110^{+0.024}_{-0.031}$ & & $0.112^{+0.014}_{-0.011}$ & \\
\hline
\end{tabular}
\vspace*{-1mm}
\caption{An overview of our statistics at different masses and in different
topological sectors. Below we display our results for
the leading non-zero eigenvalue $\lambda$
of $D_{\rm ovHF}^{(0)}$ (with jack-knife errors).
The values of $\Sigma (m)$ were obtained based on the ratio between 
$\langle \lambda \rangle $ in the topological sectors with $| \nu | =0 $ 
and $1$.}
\vspace*{-3mm}
\label{Sigmatab}
\end{table}

At last we address the {\em chiral condensate}; it has been studied 
in the 1-flavour Schwinger model with quenched configurations in 
Refs.\ \cite{DHN}.
For our case of two degenerate flavours,\footnote{For related work 
in QCD with dynamical overlap
fermions and 1 or 2 flavours, see Refs.\ \cite{SigmaQCD}.}
analytic predictions were obtained for \
$m \ll 1/\sqrt{\beta}$ \ at low energy \cite{HHI,Smilga}.
This is realised in our settings, perhaps
up to the case $m=0.24$. In particular, Ref.\ \cite{Smilga}
predicts $\Sigma (m) = - \langle \bar \psi \psi \rangle
\simeq 0.388 (m / \beta )^{1/3}$ 
based on bosonisation, while Ref.\ \cite{HHI} arrived at a slightly 
lower coefficient $\approx 0.37$. Both predictions are marked in
Fig.\ \ref{Sigmafig} (below, on the right).

\FIGURE{
\includegraphics[angle=270,width=.44\linewidth]{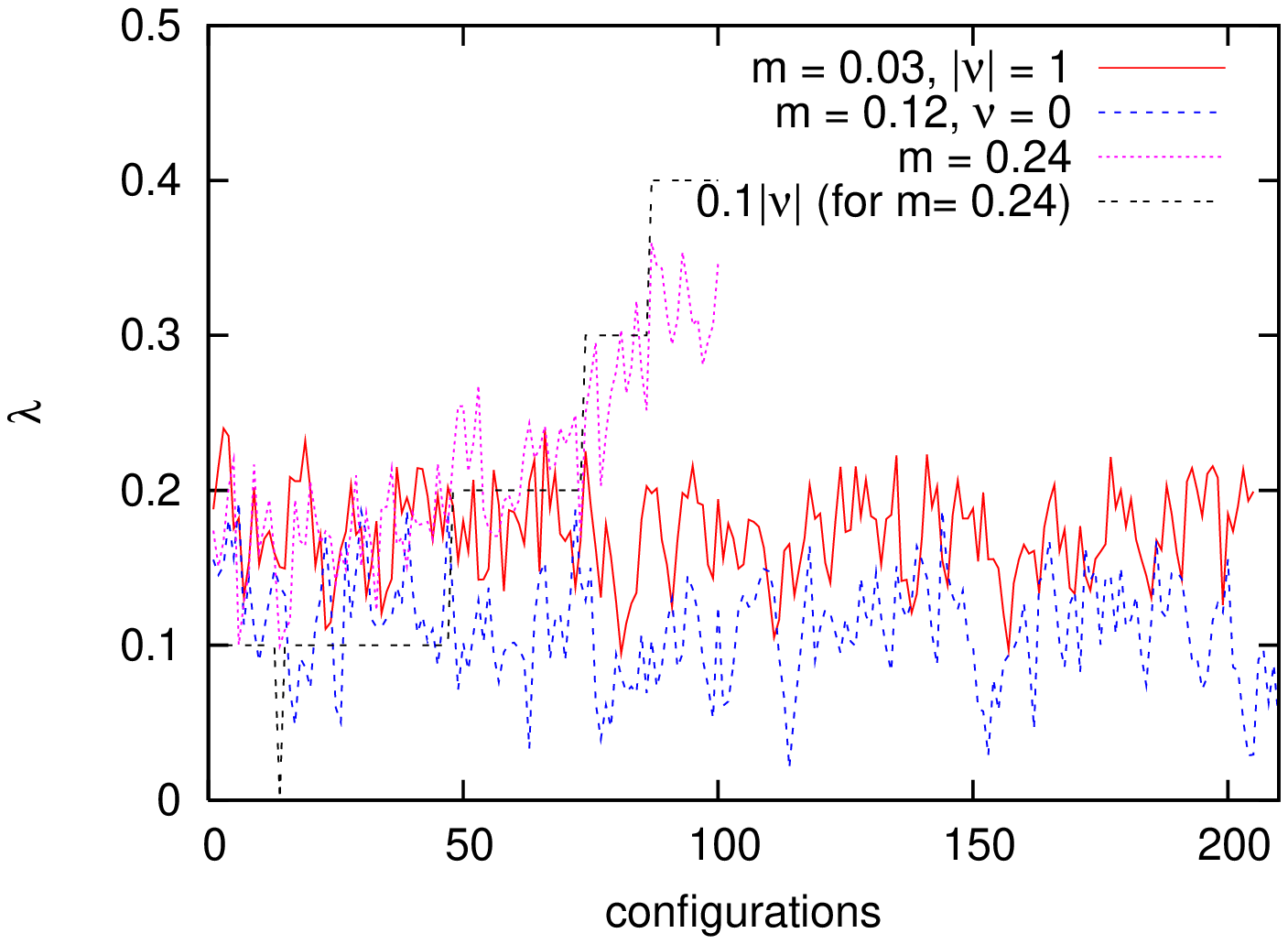} 
\includegraphics[angle=270,width=.44\linewidth]{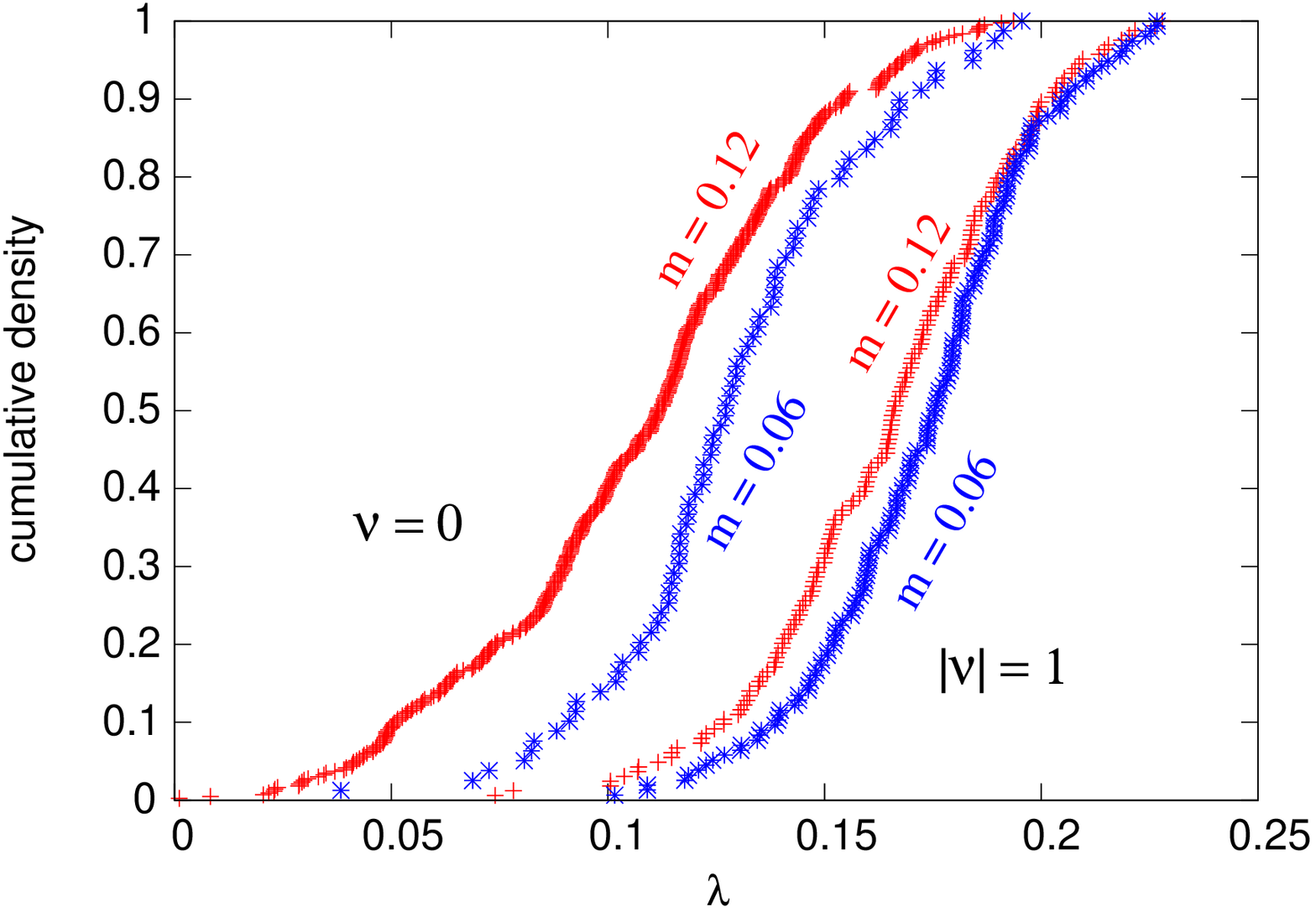} 
\includegraphics[angle=270,width=.44\linewidth]{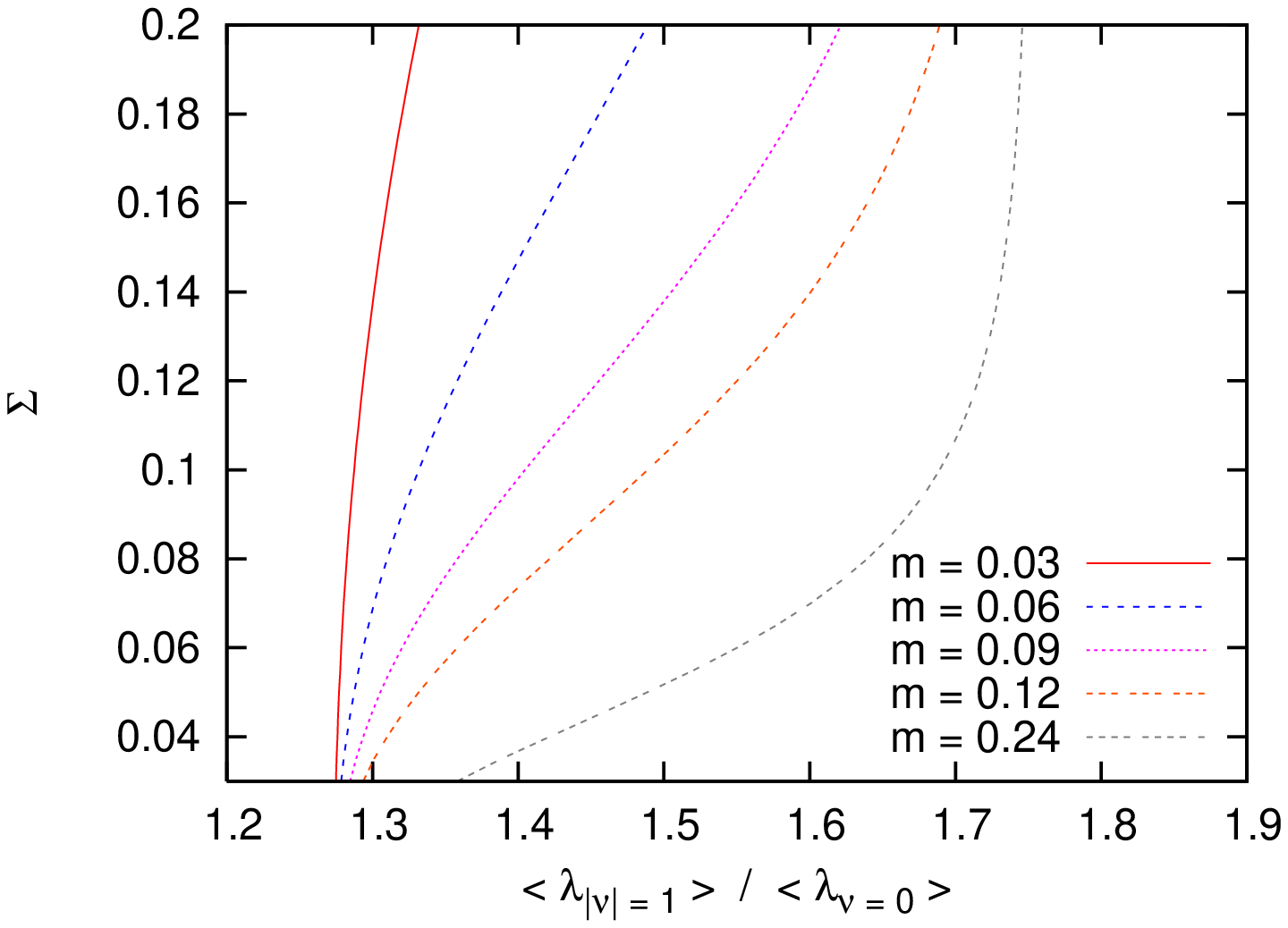} 
\includegraphics[angle=270,width=.44\linewidth]{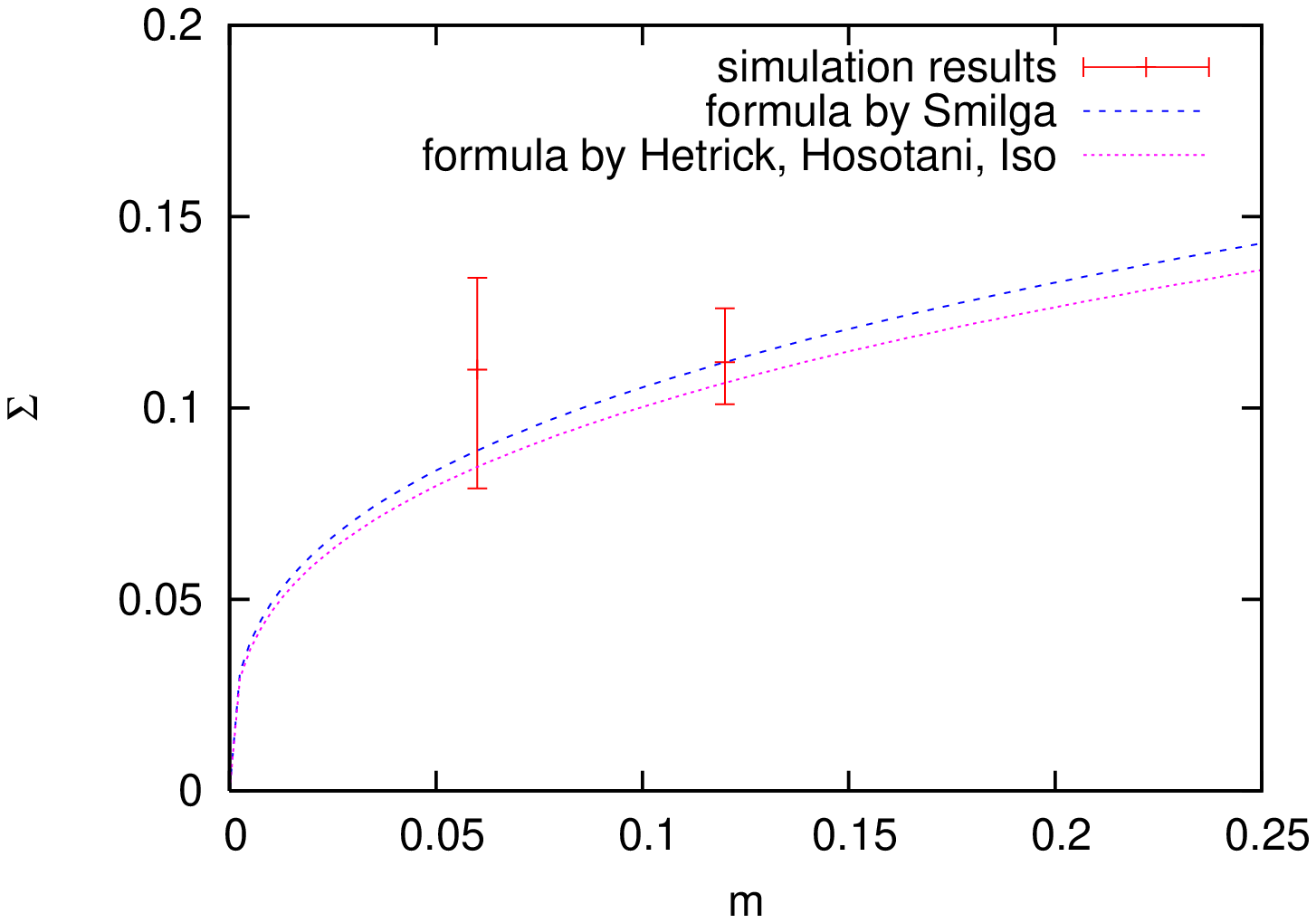}
\caption{Above: histories (left) and cumulative
densities (right) of the leading non-zero Dirac 
eigenvalue $\lambda$ at different masses and in different topological sectors.
Below on the left: the chiral condensate $\Sigma$ as a function of the ratio
$\langle \lambda_{| \nu | =1} \rangle / \langle \lambda_{\nu =0} \rangle$,
according to Random Matrix Theory \cite{WGW}.
Below on the right we show the $\Sigma$ values that we
measured at two masses, cf.\ Table \ref{Sigmatab}, and
the predictions of Refs.\ \cite{HHI,Smilga}.}
\label{Sigmafig}
}

\vspace*{-1cm} \noindent
Fig.\ \ref{Sigmafig} (above) shows
examples for HMC histories (left, illustrating the level of de-correlation),
and cumulative densities (right), of the leading non-zero
Dirac eigenvalues $\lambda$.\footnote{We used the eigenvalues of
$D_{\rm ovHF}^{(0)}$, stereographically projected on the imaginary axis,
where we take the absolute value. This treatment worked well also in quenched
QCD \cite{BJS}.}
For our evaluation of the chiral condensate we made use of a formula 
given in Ref.\ \cite{WGW} (for the $\epsilon$-regime), which expresses
$\Sigma (m)$ as a function of the ratio between $\langle \lambda \rangle$
in the sectors with topological charge $| \nu | = 0$ and $1$.
For the masses considered, these functions are 
plotted in Fig.\ \ref{Sigmafig} (below, left).
We read off $\Sigma$ for the eigenvalue ratios 
that we measured at $m = 0.06$ and $m = 0.12$, see Table \ref{Sigmatab}.
Fig.\ \ref{Sigmafig} (below, right) illustrates our results, which agree
with the predictions of Ref.\ \cite{HHI,Smilga} within the errors.


{\bf Conclusions:} \ 
We tested a force preconditioned HMC algorithm for the simulation of 
dynamical overlap fermions. It is applicable for the overlap-HF, but not
for the standard overlap fermion, since it is designed for the
case that the overlap kernel is similar to the overlap operator.
In the 2-flavour Schwinger model
we obtained a useful acceptance rate and a decent precision of the
reversibility. A high level of locality is confirmed. 
We measured the chiral condensate at two
fermion masses, and we obtained values for $\Sigma (m)$ consistent with
analytic predictions at low energy. 

Since the way to evaluate $\Sigma$ presented here
leads to relatively large errors --- the slopes of the functions
shown in Fig.\ \ref{Sigmafig} (below, on the left) tend to be steep --- we 
are now going to consider different methods for this purpose, along with an
enlarged statistics. \\  

\vspace*{-3mm}

\noindent
{\small\it We thank M.\ Hasenbusch for helpful advice,
and S. D\"{u}rr, A.\ Kennedy and J.\ Verbaarschot for 
comments.
J.V.\ was supported by the ``Deutsche Forschungsgemeinschaft'' 
(DFG). The computations were performed on the IBM p690 
clusters of the ``Norddeutscher Verbund f\"ur Hoch- und 
H\"ochstleistungsrechnen'' (HLRN).}

\end{document}